\documentclass[aps,preprint]{revtex4}%
\usepackage{amsfonts}
\usepackage{amsmath}
\usepackage{amssymb}
\usepackage{graphicx}%
\begin{document}
\title{\textbf{Ferromagnetic/superconducting proximity effect in La}$_{0.7}%
$\textbf{Ca}$_{0.3}$\textbf{MnO}$_{3}$\textbf{\ /YBa}$_{2}$\textbf{Cu}$_{3}%
$\textbf{O}$_{7-\delta}$\textbf{\ superlattices}}
\author{Z. Sefrioui$^{a}$, D Arias$^{a,\text{ }\&}$, V. Pe\~{n}a$^{a}$, J.E.
Villegas$^{b}$ , M. Varela$^{a}$,  P. Prieto$^{c}$, C. Le\'{o}n$^{a}$ J. L.
Martinez$^{d}$ and J Santamaria$^{a,\text{ }\#}$}
\affiliation{$^{a)}$\textit{GFMC, Dpto. Fisica Aplicada III, Universidad Complutense de
}Madrid, 28040 Madrid, Spain$^{b)}$\textit{Dpto. Fisica de los Materiales,
Universidad Complutense de }Madrid, 28040 Madrid, Spain$^{c)}$%
\textit{\ Departamento de F\'{\i}sica. Universidad del Valle A. A. 25360
}Cali. Colombia $^{d)}$\textit{Instituto de Ciencia de Materiales de Madrid,
CSIC, }28049 \textit{Cantoblanco}, Spain}
\date{}

\begin{abstract}
We study the interplay between magnetism and superconductivity in high quality
YBa$_{2}$Cu$_{3}$O$_{7}$(YBCO) / La$_{0.7}$Ca$_{0.3}$MnO$_{3}$(LCMO)
superlattices. We find evidence for the YBCO superconductivity depression in
presence of the LCMO layers. We show that due to its short coherence length
superconductivity survives in the YBCO down to much smaller thickness in
presence of the magnetic layer than in low Tc superconductors. We also find
that for a fixed thickness of the superconducting layer, superconductivity is
depressed over a thickness interval of the magnetic layer in the 100 nm range.
This is a much longer length scale than that predicted by the theory of
ferromagnetic/superconducting proximity effect.

\end{abstract}
\maketitle

\newpage

Ferromagnetic (F) /Superconducting (S) proximity effect has been a subject of
intense research in recent years due to the rich variety of phenomena
resulting from the competition between both long range orderings. In this
context F/S superlattices have been extensively used in the past because they
offer the possibility of tailoring individual thicknesses or modulation length
to match characteristic length scales governing ferromagnetism,
superconductivity or their interaction. Most research in this field has
involved single element or alloy based metallic superlattices [1-9]. The
extension of concepts of the F/S proximity effect to the high T$_{c}$
superconductors (HTS) or colossal magnetoresistance (CMR) oxides is of primary
interest since peculiarities like the short superconducting coherence length
and full spin polarization could open the door to interesting new effects.
Although there has been recently a theoretical effort to examine the F/S
interface in oxides [10], to the best of our knowledge, experimental results
on F/S proximity effect are lacking in the literature. In this paper we
examine the interplay between magnetism and superconductivity in YBa$_{2}%
$Cu$_{3}$O$_{7}$(YBCO) / La$_{0.7}$Ca$_{0.3}$MnO$_{3}$(LCMO) superlattices and
provide evidence for superconductivity depression due to the presence of
magnetic layers. YBCO and LCMO have oxide perovskite structure with very
similar in-plane lattice parameters, which allows the growth of superlattices
with sharp interfaces, thus strongly reducing extrinsic (structural) effects
which otherwise could obscure the F/S interplay.

At the F/S interface, Cooper pairs entering the ferromagnet from the
superconductor experience the exchange interaction, which favors one of the
spin orientations. This causes the superconducting order parameter to decay in
the F layer faster than in a normal metal, within a length scale $\xi_{F}
$=$\hbar$v$_{F}$/ $\Delta$E$_{ex}$ (where v$_{F}$ is the Fermi velocity and
$\Delta$E$_{ex}$ the exchange splitting). In single element or alloy
ferromagnets, for typical values of $\Delta$E$_{ex}$ =1eV and v$_{F}$ of
10$^{8}$ cm/s, $\xi_{F}$ is of the order of 1 nm [3] which is shorter than the
superconducting coherence length of the low temperature superconductors,
(usually larger than 10 nm). Superconductivity is also depressed in the S
layer within a characteristic length scale, $\xi_{S}$, given by ($\hbar$%
D$_{S}$/ k$_{B}$T$_{c}$)$^{0.5}$[8], where D$_{S}$ is the electron diffusion
coefficient for the superconductor. $\xi_{S}$ is of the order of the
superconducting coherence length. Thus, the critical temperature of the
superlattice can be much smaller than that corresponding to the bulk
superconductor. For F/S superlattices, this results in a critical thickness of
the superconducting layer, d$_{cr}^{S}$, below which superconductivity is
suppressed. The thickness of the F layers tunes the coupling between the S
layers, yielding a critical F layer thickness d$_{cr}^{F}$($\sim$2$\xi_{F}$),
above which superconducting critical temperature should become independent of
the thickness of the magnetic layer (decoupled S layers). Many experimental
data on metallic (single element) samples have been analyzed using the
theoretical approach by Radovic et al. [8] based on Usadel equations [11] and
in the De Gennes -Werthammer boundary conditions [12]. Within the frame of
this theory, quite exotic phenomena have been predicted and experimentally
observed [13-17] for thin magnetic layers, such that the superconducting
layers are coupled (d$^{F}<$ d$_{cr}^{F}\sim$2$\xi_{F}$). A $\pi$ phase shift
of the order parameter between the superconducting layers yields an
oscillating order parameter along the direction normal to the interface, which
also gives rise to oscillating dependence on F layer thickness of T$_{c}$,
critical currents and fields.

In our system, due to the short YBCO coherence length (0.1-0.3 nm), S layers
are expected to sustain superconductivity down to much thinner thickness than
in the case of conventional (low temperature) superconductors. On the other
hand, the F material LCMO shows a large exchange splitting (3 eV) and
relatively small bandwidth, giving rise to a fully spin polarized conduction
band [18], which may suppress superconducting proximity effect into LCMO over
very short length scales (small $\xi_{F}$). The high quality LCMO/YBCO
superlattices used in this work allow us to investigate both issues.

Samples were grown in a high pressure (3.4 mbar) pure oxygen sputtering system
at high temperatures (900
${{}^o}$%
C). Individual YBCO films on STO (100) were epitaxial with T$_{c}$ of 90 K and
transition widths smaller than 0.5 K. Growth conditions, optimized for the
YBCO, yielded LCMO single films with a ferromagnetic transition temperature
T$_{CM}$= 200 K, and a saturation magnetization M$_{S}$= 400 emu/cm$^{3}$,
close to the bulk value. Two sets of samples were grown for this study:
superlattices with fixed YBCO thickness (5 unit cells per bilayer) and
changing LCMO thickness between 1 and 100 unit cells (set A) up to a total
thickness of 150 nm; and superlattices with fixed LCMO thickness (15 unit
cells) and changing YBCO thickness from 1 to 12 unit cells (set B). Samples
were checked for the simultaneous presence of magnetism and superconductivity
[19,20] by transport (resistivity) and susceptibility (SQUID) measurements.

Figure 1 shows x-ray diffraction (XRD) patterns of a sample with very thin
YBCO (1 unit cell) and of a sample with very thin manganite (3 unit cells).
The corresponding TEM cross section views of the same superlattices, obtained
in a Philips CM200 microscope operated at 200 kV are also shown. Clear
superlattice Bragg peaks and satellites can be observed, which together with
the flat interfaces of the TEM pictures show a high degree of structural
order. XRD patterns were checked for the presence of interface disorder using
the SUPREX 9.0 refinement software [21]. The calculated spectra, which are
really close to the experimental data, only include step disorder at the
interface consisting of 0.5-0.7 manganite unit cells. Refinements were
consistent with the absence of interdiffusion. In fact, the incorporation of
small amounts ($<$10\%) of La or Ca into Y sites considerably deteriorated the
agreement between experimental and calculated spectra. We also found no
indications of epitaxial mismatch strain (x ray refinement did not show
changes in the lattice parameters along the c direction) as expected from the
small lattice mismatch between YBCO and LCMO.

Samples were magnetic down to the smallest LCMO layer thickness. The inset of
Figure 2 shows hysteresis loops measured at 90 K (above the superconducting
transition) and with magnetic fields parallel to the layers, of samples of set
A with 5, 12 and 18 unit cells thick LCMO layers. A systematic reduction of
the magnetization with LCMO thickness is observed (see main panel of figure
2), which has been reported for ultrathin LCMO layers grown on various
substrates [22, 23]. The inset of figure 3 shows resistance curves of
representative samples of set A (constant YBCO thickness of 5 unit cells,
changing LCMO layer thickness). While the superlattices with thinner LCMO
layers show superconducting critical temperatures close to bulk YBCO values, a
systematic depression of the critical temperature is observed for samples with
LCMO layers thicker than 5 unit cells. The metal insulator transition
associated to the ferromagnetic transition can be observed in the thicker LCMO
layers. Main panel of figure 3 shows the evolution of T$_{c}$ with LCMO layer
thickness. It is important to notice that T$_{c}$ keeps on decreasing over a
very large LCMO thickness interval. \textbf{ }The inset of figure 4 shows
resistance curves for a series of superlattices with increasing YBCO thickness
(set B). It can be observed that the superconductivity is completely
suppressed for YBCO layer thickness of 1 and 2 unit cells. For larger YBCO
layer thickness, however, T$_{c}$ displays a monotonic increase up to a value
of 85 K (close to that of thick single films) for N=12. Main panel of figure 4
shows the evolution of T$_{c}$ with YBCO layer thickness.

A depression of the critical temperature of ultrathin YBCO layers (1-5 unit
cells) has been also observed in presence of non magnetic spacers of fixed
thickness [24]. However we show here that superconductivity is further
depressed in presence of the magnetic layers. While 1 unit cell YBCO layer is
still superconducting in presence of 5 unit cells thick PrBa$_{2}$Cu$_{3}%
$O$_{7}$(PBCO) layers in YBCO / PBCO superlattices, with a T$_{c}$ of 30 K, 1
unit cell of YBCO in presence of the same thickness (15 unit cells) of LCMO is
non superconducting. In addition, 5 unit cells of YBCO in presence of PBCO
have already the bulk T$_{c}$, whilst in presence of LCMO a reduced T$_{c}$ of
50 K is observed. Other extrinsic factors for the depression of T$_{c}$ like
deficient oxygenation of the YBCO through the manganite layers can be ruled
out since the thickest (above 9 unit cells) YBCO layers almost completely
recover the bulk critical temperature (figure 4). Moreover, the fact that
quite thick LCMO layers are necessary to reduce the T$_{c}$ of 5 YBCO unit
cells supports that the changes of T$_{c}$ are not due to interdiffusion: if
T$_{c}$ decrease were due to interdiffusion one would expect the greatest
effect for the first (few) LCMO unit cells. Tc depression in presence of the
magnetic layers, thus, indicates the interaction between magnetism and
sperconductivity. In fact there is an additional result pointing in this
direction. We have found a clear correlation between the critical temperature
and the magnetic moment of LCMO. Figure 5 shows that enhanced magnetization of
the thicker LCMO layers results in lower Tc values.

We discuss now the possibility of F/S proximity effect in these samples. Due
to the F/S proximity effect, the superconducting order parameter within the S
layer decays with a characteristic length scale $\xi_{S}$, given by ($\hbar
$D$_{S}$/ k$_{B}$T$_{c}$)$^{0.5}$[8]. An estimate using the resistivity of the
YBCO normal to the CuO planes yields $\xi_{S}$ =0.6 nm, relatively close to
the superconducting coherence length, $\xi_{c}$, (0.1-0.3 nm). In view of
figure 4, superconductivity is suppressed for a critical thickness d$_{cr}%
^{S}\approx$ 3 nm (between 2 and 3 unit cells), i.e. d$_{cr}^{S}$ /$\xi
_{S}\approx$ 5 roughly. It is interesting to note that this value of
d$_{cr}^{S}$ is considerably smaller than those found in metallic
superlattices with low T$_{c}$ superconductors for similar thickness of the
magnetic spacer and also for values of the magnetization which are not very
different from that of the 15 manganite unit cells. For example Aarts et al.
[2] reported d$_{cr}^{S}$ = 25 nm for [V/V$_{0.34}$ Fe$_{0.66}$] superlattices
and Lazar et al. [6] found d$_{cr}^{S}$ = 70 nm in Fe/Pb/Fe trilayers. I.e.,
much shorter coherence length of YBCO compared to low T$_{c}$ superconductors
allows superconductivity to exist down to quite small thicknesses in presence
of magnetic layers. On the other hand superconductivity induced within the F
layer decays with a length scale $\xi_{F}$= $\hbar$v$_{F}$/$\Delta$E$_{ex}$.
Given the large exchange splitting of the LCMO (3 eV) and a Fermi velocity for
the majority band of 7.4 10 $^{7}$ cm/s [18], the former expression yields
very small values for $\xi_{F}$ of about 0.2 nm. Therefore, the large exchange
splitting of the manganite strongly unfavors the superconducting proximity
effect. In experiments changing the thickness of the magnetic layer (d$_{F}$)
with fixed thickness of the superconducting layer one expects, according to
previous theoretical approaches [8], that Tc is depressed for magnetic layer
thickness smaller than $\xi_{F}$, and that Tc saturates at a d$_{F}$
independent value for larger thickness of the magnetic layer (d$_{F}>\xi_{F}%
)$. Keeping in mind the short values estimated for $\xi_{F}$, this is at
variance to what is observed in figure 3. We have found that Tc is still
changing for thicknesses of the magnetic layer which are more than two orders
of magnitude larger than the estimated value for $\xi_{F}$. In fact more than
50 manganite unit cells (19 nm) are necessary to suppress the
superconductivity of 5 YBCO unit cells, suggesting that a much longer length
scale than $\xi_{F}$ is ruling the superconductivity suppression in these
oxide systems. The reduced magnetic moment of the thinnest magnetic layers
shown in figure 5 might be invoked to propose an explanation for the long
length scale for superconductivity suppression into the ferromagnet. The
exchange splitting $\Delta$E$_{ex}$ is the energy difference between electrons
at the Fermi level, with spins parallel or antiparallel to the magnetization.
$\Delta$E$_{ex}$ is connected to the magnetic moment $\mu_{F}$ through
$\Delta$E$_{ex}$=I$_{eff}\quad\mu_{F}$, where I$_{eff}$ is an effective
exchange integral. Thus one expects that $\xi_{F}$ can be enlarged by the low
magnetic moment. In fact 1/$\xi_{F}$ has been shown to increase linearly with
magnetic moment of V-Fe alloys in V/ V-Fe F/S multilayers [2]. In our LCMO
layers $\mu_{F}$ is reduced by more than 20 times (respect to bulk values) for
the thinnest layers and by a factor of 2 for the thick 50 unit cells LCMO
layers. This could explain an apparent enlargement of $\xi_{F}$ for the
thinnest LCMO layers, but still does not explain the decrease of Tc of the 5
unit cells of YBCO of figure 3 for the thickest 20-50 LCMO layers.

An additional complication trying to explain the figure 3 with the F/S
proximity effect is related to interface transparency. From figure 3 we find
an apparent distance into the ferromagnet (over which Tc is suppressed) which
is orders of magnitude longer than the theoretical estimates. Reduced
interface transparency due to interface disorder would shorten this distance
contrary to what is observed [25]. Moreover, it has been proposed that at the
interface with thick half metallic ferromagnets pairs will experience complete
reflection due to the energy separation between the bottom of the minority
subband and the Fermi level [2]. This is equivalent to a vanishing interface
transparency in the formalism of the F/S proximity effect [2] . It is clear
then that the behavior of figure 3 cannot be explained by the conventional
theory of the F/S proximity effect. However, we want to remark in this respect
that there is not a theory for the F/S proximity effect for fully spin
polarized ferromagnets.

We speculate that the injection of spin polarized carriers from the LCMO into
the YBCO \ may add a new source of superconductivity depression: pair breaking
by spin polarized carriers. This mechanism has been theoretically analyzed
before [26] and recently observed in manganite/HTS junctions as a depression
of the critical current with the injected spin polarized current [27,
28].\textbf{ }The injection of spin polarized carriers over the
superconducting gap depresses the order parameter monotonically with
increasing the quasiparticle density. In the simplest picture this depression
can be accounted for [29] by $\frac{{\Delta\left(  {n_{qp}}\right)  }}%
{{\Delta\left(  {0}\right)  }}\cong1-\frac{{n_{qp}}}{{2\Delta\left(
{0}\right)  N\left(  {0}\right)  }}$where $\Delta$(n$_{qp}$) is the depressed
energy gap by the quasiparticle density n$_{qp}$, $\Delta$(0) is the zero
temperature energy gap and N(0) the density of states at the Fermi level. At
low temperatures where the thermally induced quasiparticle density is small,
recombination of injected spin polarized carriers requires spin flip
scattering what considerably increases their diffusion time. This pair
breaking effect extends over the spin diffusion length (\textit{l}$_{S}$) into
the superconductor which can be very long; for example, a value of the order
of 1 cm has been reported for Al [30]. An estimate of \textit{l}$_{S}$ in YBCO
can be obtained following ref. 27, using the relation \textit{l}$_{S}%
$=(\textit{l}$_{0}$ v$_{F}$ $\tau_{S}$)$^{0.5}$[28], where $\tau_{S}$ is the
spin polarized quasiparticle diffusion time, v$_{F}$ is the Fermi velocity and
\textit{l}$_{0}$ the electron mean free path. Assuming a value of $\tau_{S}$
=10$^{-13}$ s [28], v$_{F}$ = 10$^{7}$ cm/s and that the electron mean free
path is limited by YBCO layer thickness \textit{l}$_{0}$= 6 nm,\textit{\ l}%
$_{S}$ can be as long as 8 nm. This length scale compares favorably with the
thickness of the superconducting layer (6 nm), and suggests that pair breaking
by injected spin polarized carriers could play a role in the superconductivity
suppression in our F (CMR)/ S (HTS) superlattices. Further work wil be
necessary to highlight this point.

In summary, we have provided evidence for superconductivity depression by the
presence of the magnetic layer in LCMO/YBCO superlattices. A structural study
using TEM and x ray refinement has been used to discard extrinsic effects like
interface disorder or interdiffusion. We have found that YBCO
superconductivity is depressed in presence of manganite layers with a
characteristic length scale much longer than that predicted by the existing
theories of the F/S proximity effect. This result should provide an avenue for
future theoretical studies of the F/S proximity effect in presence of spin
polarized ferromagnets.

\textbf{Acknowledgments}

Work supported by CICYT MAT2000-1468, CAM 07N/0008/2001 and Fundaci\'{o}n
Ram\'{o}n Areces. We thank R. Escudero, J. Fontcuberta, J. M. de Teresa, C. Sa
de Melo, I. K. Schuller, J. L. Vicent, and V. Vlasko-Vlasov for useful conversations.

\newpage REFERENCES

$^{\&} $\textit{On leave from Universidad del Quindio, Armenia, Colombia.}

$^{\#}$\textit{\ Corresponding author: J. Santamaria, email: }%
jacsan@fis.ucm.es\textit{.}

[1]C. Uher, R. Clarke, G.G. Zheng, and I.K. Schuller, Phys. Rev. B\textbf{30},
453 (1984).

[2] J. Aarts, J. M. E. Geers, E. Br\"{u}ck, A. A. Golubov, and R. Coehorn,
Phys. Rev. B \textbf{56}, 2779 (1997).

[3] Th. M\"{u}hge et al., Phys. Rev. B\textbf{57}, 5071 (1998).

[4] S. Kaneko et al., Phys. Rev B\textbf{58}, 8229 (1998).

[5] G. Verbanck et al., Phys. Rev. B\textbf{57}, 6029 (1998).

[6] L. Lazar et al., Phys. Rev. B\textbf{61}, 3711 (2000).

[7] F.Y. Ogrin, S.L. Lee, A.D. Hilier, A. Mitchell and T.-H. Shen, Phys. Rev.
B\textbf{62}, 6021 (2000).

[8] Z. Radovic, L. Dobrosavljevic-Grujic, A. I. Buzdin and J. R. Clem, Phys.
Rev. B \textbf{38}, 2388 (1988). Z. Radovic et al., Phys. Rev. B \textbf{44},
759 (1991).

[9] M. Velez et al., Phys. Rev. B \textbf{59}, 14659 (1999).

[10] C. A. R. S\'{a} de Melo, Phys. Rev. Lett \textbf{79,} 1933 (1997);
\textit{ibid}. Phys. Rev. B\textbf{62}, 12303 (2000).

[11] K. Usadel, Phys. Rev. Lett \textbf{25,} 507 (1970).

[12] N.R. Werthamer, Phys. Rev. \textbf{132}, 2440 (1963); P.G. de Gennes,
Rev. Mod. Phys. \textbf{36}, 225 (1964).

[13] Jhon Q. Xiao and C.L. Chien, Phys. Rev. Lett.\textbf{76}, 1727(1996).

[14] Th. M\"{u}hge et al., Phys. Rev. Lett. \textbf{77}, 1857 (1996).

[15] L. V. Mercaldo et al., Phys. Rev. B\textbf{53}, 14040 (1996).

[16] J.S. Jiang, Dragomir Davidovic, Daniel H. Reich, and C.L. Chien, Phys.
Rev. B\textbf{54}, 6119 (1996).

[17] V.V. Ryazanov et al., Phys. Rev. Lett. \textbf{86}, 2427 (2001).

[18] Warren E. Pickett and David J. Singh, Phys. Rev. \textbf{B}53, 1146 (1996).

[19] G. Jakob, V. V. Moshchalkov, and Y. Buynseraede, Appl. Phys. Lett.
\textbf{66}, 2564 (1995).

[20] P. Prieto et al., J. Appl. Phys. \textbf{89}, 8026 (2001); H.-U.
Habermeier et al., Physica C \textbf{354}, 298 (2001).

[21] I. K. Schuller, Phys. Rev. Lett. \textbf{44}, 1597 (1980); W Sevenhans et
al., Phys. Rev. B \textbf{34}, 5955 (1986); E. E Fullerton, I. K. Schuller, H.
Vanderstraeten and Y. Bruynseraede, \textit{ibid.}. B \textbf{45}, 9292
(1992); D. M. Kelly, E. E Fullerton, J. Santamaria and I. K. Schuller, Scripta
Met. Mat. \textbf{33}, 1603 (1995).

[22] M.Bibes et al., Phys. Rev. Lett. \textbf{87}, 067210 (2001).

[23] M. Ziese, H.C. Semmelhack, K.H. Han, S.P. Sena and H.J. Blythe, J. Appl.
Phys. \textbf{91}, 9930 (2002).

[24] M. Varela et al., Phys. Rev. Lett. \textbf{83}, 3936 (1999).

[25]\textbf{ }I. Baladie and A. Buzdin cond-mat/0209466.

[26] A. G. Aronov, Sov. Phys. JETP \textbf{19}, 1228 (1964).

[27] V. A. Vas%
\'{}%
ko et al., Phys. Rev. Lett. \textbf{78,} 1134 (1997). A. M. Goldman et al., J.
Magn. Magn. Mater. \textbf{200}, 69 (1999).

[28] N. C. Yeh et al., Phys. Rev. B \textbf{60}, 10522 (1999).

[29] W. H. Parker, Phys. Rev. B12, 3667, (1995)

[30] Mark Johnson and R.H. Silsbee, Phys. Rev. Lett. \textbf{17}, 1790 (1985).

\newpage FIGURE CAPTIONS

\bigskip

\textbf{Figure 1: (a)} TEM cross section view of a [LCMO (3 u.c.)/YBCO (5
u.c.)] superlattice.

\textbf{(b)} X- Ray difracction pattern and SUPREX calculated spectra of
sample [LCMO (3 u.c.)/YBCO (5 u.c.)]

\textbf{(c)} TEM cross section view of a [LCMO (15 u.c.)/YBCO (1 u.c.)] superlattice.

\textbf{(d)} X- Ray difracction pattern and SUPREX calculated spectra of
sample [LCMO (15 u.c.)/YBCO (1 u.c.)]

\bigskip

\textbf{Figure 2:} Saturation magnetization vs. LCMO thickness for
superlattices [LCMO (N$_{M}$ u.c.)/YBCO (5 u.c.)]. Line is a guide to the eye.
Inset shows hysteresis loops at T=90 K for samples with N$_{M}$=5 (squares),
N$_{M}$ =12 (triangles) and N$_{M}$ =18 (circles) unit cells.

\bigskip

\textbf{Figure 3:} T$_{c}$ vs. LCMO thickness for [LCMO (N$_{M}$ u.c.)/YBCO (5
u.c.)] superlattices. Inset: Resistance vs. temperature curves for N$_{M}$=3,
9,15,60,90 unit cells (from bottom to top).

\bigskip

\textbf{Figure 4:} T$_{c}$ vs. YBCO thickness for [LCMO (5 u.c.)/YBCO (N$_{S}$
u.c.)] superlattices. Inset: Resistance vs. temperature curves for N$_{S}%
$=1,2,3,4,5,6,8,12 (from top to bottom).

\bigskip

\textbf{Figure 5:} T$_{c}$ vs. saturation magnetization for [LCMO (N$_{M}$
u.c.)/YBCO (5 u.c.)] superlattices.
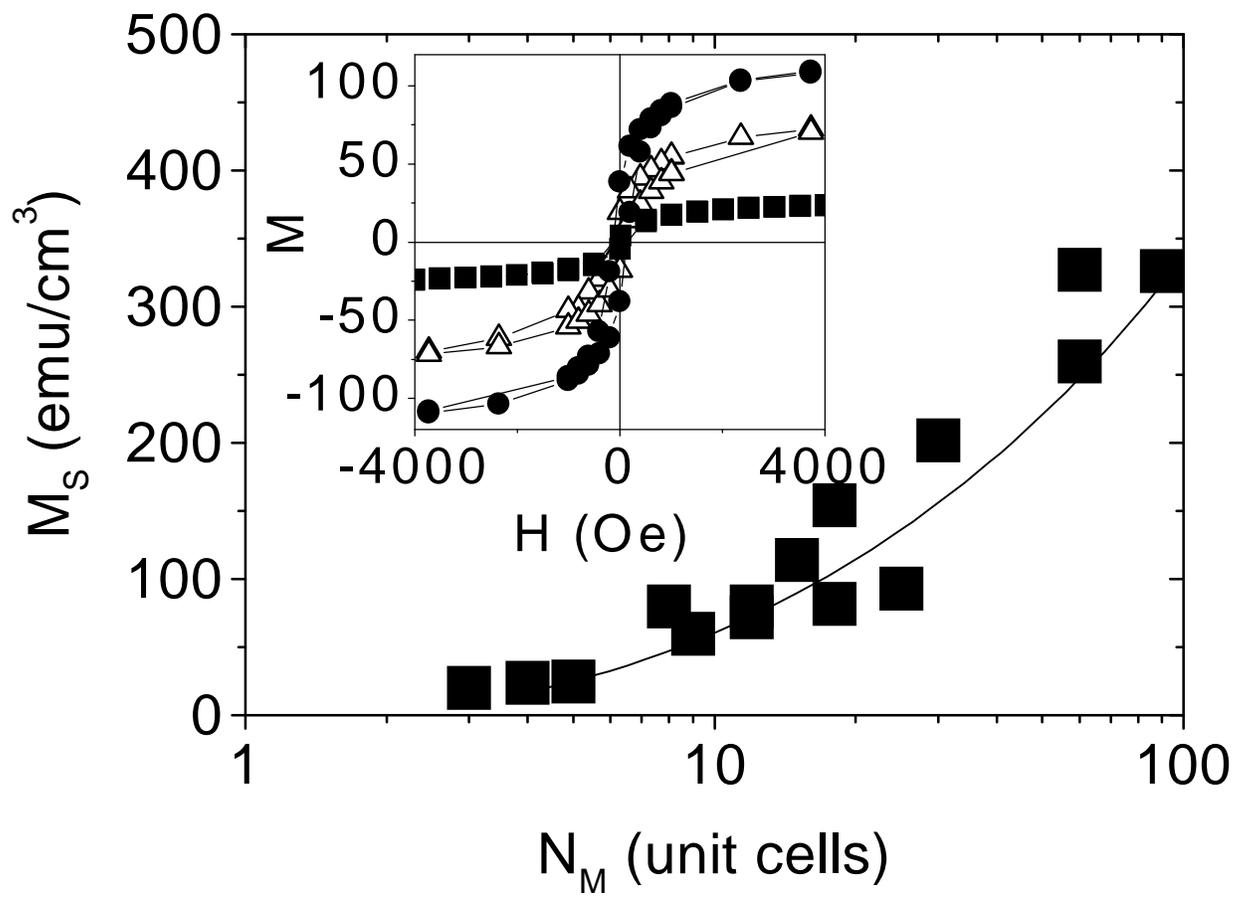
\end{document}